\documentclass[showpacs,amsmath,amssymb,aps,pre]{revtex4}
\usepackage{graphicx}

\begin{document}
\title{Study Of The Wealth Inequality In The Minority Game}
\author{K. H. Ho, F. K. Chow and H. F. Chau}
\affiliation{Department of Physics, University of Hong Kong, Pokfulam Road,
Hong Kong}
\date{\today}
\begin{abstract}
To demonstrate the usefulness of physical approaches for the study of 
realistic economic systems, we investigate the inequality of players' wealth 
in one of the most extensively studied econophysical models, namely, the 
minority game (MG). We gauge the wealth inequality of players in the MG by a 
well-known measure in economics known as the modified Gini index. From our numerical 
results, we conclude that the wealth inequality in the MG is very severe near the 
point of maximum cooperation among players, where the diversity of the strategy 
space is approximately equal to the number of strategies at play. In other 
words, the optimal cooperation between players comes hand in hand with 
severe wealth inequality. We also show that our numerical results 
in the asymmetric phase of the MG can be reproduced semi-analytically using a
replica method.
\end{abstract}


\maketitle

\section{Introduction}
Econophysics is the study of economic systems by employing methods and tools
developed in physics. Up to now, many economists have been worrying that
econophysicists are just reinventing the wheel; while many physicists are 
studying properties of toy economic models that are not directly relevant to 
economics~\cite{Fin}. In this paper, we investigate the inequality 
of wealth in a simple-minded econophysical model known as the minority game (MG) using
the so-called replica trick~\cite{spin1,spin2,spin3}. By doing so, we hope to 
make a small step forward in the application of physical methods when studying
real economic systems.

The MG is a simple-minded model of a complex adaptive system which 
captures the cooperative behavior of selfish players in a real market. In this
game, $N$ players have to choose one of the two possible alternatives in
each turn based only on the minority sides in the previous $M$ turns. The 
wealth of those who end up in the minority side will be increased by one while 
the wealth of the others will be reduced by one. To aid the players in making 
their choice, each of them is randomly and independently assigned $S$ 
deterministic strategies once and for all when the game begins. 
Each deterministic strategy is nothing but a map from the set of all possible 
histories (a string of the minority side of the previous $M$ turns) to the set 
of the two possible alternatives. All players make their choices
according to their current best strategies~\cite{MG1,MG2}. In the MG, the 
complexity of the system is usually indicated by the control parameter 
$\alpha \equiv 2^{M+1}/ NS$ which is the ratio of the size of the strategy 
space to the size of strategies at play~\cite{MG2, MG3, MG4}.

Clearly, the mean attendance of either choice is $N/2$ as the game is 
symmetrical for both choices. In contrast, the variance of this probability, 
which is conventionally denoted by $\sigma^2 (A)$, is highly non-trivial. It 
attains a very small value when $\alpha \approx 1$, indicating that the players are 
cooperating~\cite{MG3}.  That is why previous studies of the MG and its 
variants~\cite{EMG1,MMG1,MMG2} focus mainly on the study of $\sigma^2 (A)$.

Since the strategies are assigned once and for all to each player, it is
possible that some poorly-performing players are somehow forced to cooperate
with some well-performing peers. Therefore, it makes sense to study the 
inequality of wealth in MG in detail.

In Section~\ref{Gini_Formula}, we introduce a common method that measures 
wealth inequality in economics known as the modified Gini index. We then study 
the Gini index in the MG numerically in Section~\ref{Num_Re}. Our numerical 
simulation shows that both the maximal cooperation point and the point of 
maximum wealth inequality occur around $2^{M+1} \approx NS$. This confirms our 
suspicion that the apparent cooperation of players shown in the 
$\sigma^2 (A)$ does not tell us the complete story. In fact, we 
are able to explain the trend of a modified Gini index 
qualitatively using the crowd-anticrowd theory~\cite{CA1,CA2,CA3}. In 
particular, we find that the cooperation comes along with wealth 
inequality partially because poorly-performing players cannot change their 
strategies in the MG. In this way, we show that the crowd-anticrowd theory is 
not only able to explain $\sigma^2 (A)$, but also explains other features of 
other quantities in the MG. In Section~\ref{Replica}, we try to reproduce our 
numerically simulated Gini index in the so-called asymmetric phase using the 
replica method. we recall that one has to average over the disorder 
variables in the conventional replica method; the direct application of the
replica trick cannot provide the wealth distribution of players and thus
the Gini index of the MG. Fortunately, a careful semi-analytic application
of the replica method can be used to reproduce the Gini index qualitatively 
as a function of $\alpha$. Finally, we wrap up by giving a brief summary
of our work in Section~\ref{Con}.

\section{Gini index with negative wealth} \label{Gini_Formula}

In order to measure the inequality of wealth among players in the MG qualitatively, 
we follow our economics colleagues employing the so-called Gini index. In the 
original definition, the Gini index $G_0$ in a population is the mean 
of the absolute differences between the wealth all possible pairs of players
~\cite{Gini4}. That is to say,
\begin{equation}\label{ogini_eq}
G_0 = 1 - \frac{1}{N} \sum_{j = 1}^N \left[ 1 + 2 ( N - j ) \right] g_j.
\end{equation}
In the above equation, $N$ is the number of players in the population,
$g_j$ is the wealth earned by a player divided by the total wealth in the population. 
and the $g_j$'s are ranked in ascending order, i.e.,
$g_1 \leq g_2 \leq \ldots \leq g_N$. Clearly, $G_0$ ranges from $0$ to $1$. The larger 
$G_0$, the more serious the wealth inequality. If $G_0 = 0$, the players'
wealth is uniformly distributed. If $G_0 \simeq 1$, one of the players possesses 
the total wealth of the population and the wealth inequality is served. However, 
Eq.~(\ref{ogini_eq}) is only applicable in two cases: (1) all players have positive 
wealth; or (2) all players have negative wealth. Since players in the MG may have 
positive or negative wealth, we cannot use $G_0$, in general, to measure wealth inequality.
We employ an extension of $G_0$, introduced by Chen {\it et al.}, known as the 
modified Gini index $G$~\cite{Gini1,Gini2,Gini3}, is given by 
\begin{equation} \label{gini_eq}
	G = \frac{ \displaystyle \frac{2}{N} \displaystyle \sum_{ j = 1 }^{N} 
	j g_j - \frac{N + 1}{N} }
	{ 1 + \displaystyle \frac{2}{N} \sum_{ j = 1}^{k} j g_j + \frac{1}{N} 
	\sum_{ j = 1 }^{k} g_j 
	\left[ \frac{\sum_{j=1}^{k} g_j }{g_{k+1}} -  ( 1 + 2 k) \right] },
\end{equation} 
where $k$ is defined in such a way that $\sum_{j = 1 }^{k} 
g_j < 0$ and $\sum_{ j = 1 }^{ k + 1 } g_j > 0$. For simplicity, we refer to
the modified Gini index $G$ as the Gini index from now on. Just like the
original Gini index $G_0$, the modified Gini index $G$ measures the normalized wealth
inequality of players. Again, $G$ ranges from $0$ to $1$. The larger the value of $G$, 
the more serious the wealth inequality. When all players are equally 
wealthy, i.e., $k = 0$, the term $\sum_{j=1}^{k} j g_j$ vanishes and
$G$ becomes zero. In contrast, if the total wealth of the system is owned by a single 
player, i.e., $g_N =1$ and the term $\sum_{j=1}^{N-1} j g_j = 0$, then $G$ 
attains a value of one as $N \rightarrow \infty$. Also, $G$ is reduced to 
the original Gini index $G_0$ when all players have positive wealth or all players 
have negative wealth. Moreover, $G$ is unchanged if the wealth of each 
player is multiplied by a non-zero constant.

\section{Numerical Results and Qualitative Explanations} \label{Num_Re}

In this section, we investigate the wealth inequality 
of the players in the MG. Since we are only interested in studying the generic 
properties of the Gini index, we average the Gini index over $N_r = 500$ 
independent runs. Because $G$ measures the normalized wealth distribution 
of players rather than simply the first and second moments of this 
distribution, the time of convergence of Gini index $G$ is much longer than 
that of the variance of attendance and it differs for different initial 
configurations of the system. So we employ an adaptive scheme to check for 
system equilibration before taking any measurement. Specifically, in each run, 
we record the time series of $G$ until the absolute difference of $G$ 
between $10000$ successive steps is less than $10^{-6}$. Then, we 
obtain the equilibrated value of $G$ by using finite size scaling. Finally, we 
take $G$ to be the average over $50$ measurements each separated by $1000$ 
steps. we recall from Section~\ref{Gini_Formula} that $G$ is a measure of the 
normalized wealth distribution. From our numerical simulation, $G$ equilibrates 
logarithmically and slowly although the wealth of players is decreasing in each 
turn. We will explain the reason for convergence of $G$ in detail at 
the end of this section. We have performed numerical simulations for the cases
where players draw their strategies from full strategy space and reduced 
strategy space~\cite{MG1, MG4} respectively. The Gini indices obtained in 
these two cases are very similar. Since the analytical investigation performed in 
Section~\ref{Replica} is simpler if we focus on reduced strategy space, we 
present the numerical results based on reduced strategy space here for 
consistency.

\begin{figure}[ht]
\includegraphics*[scale = 0.6]{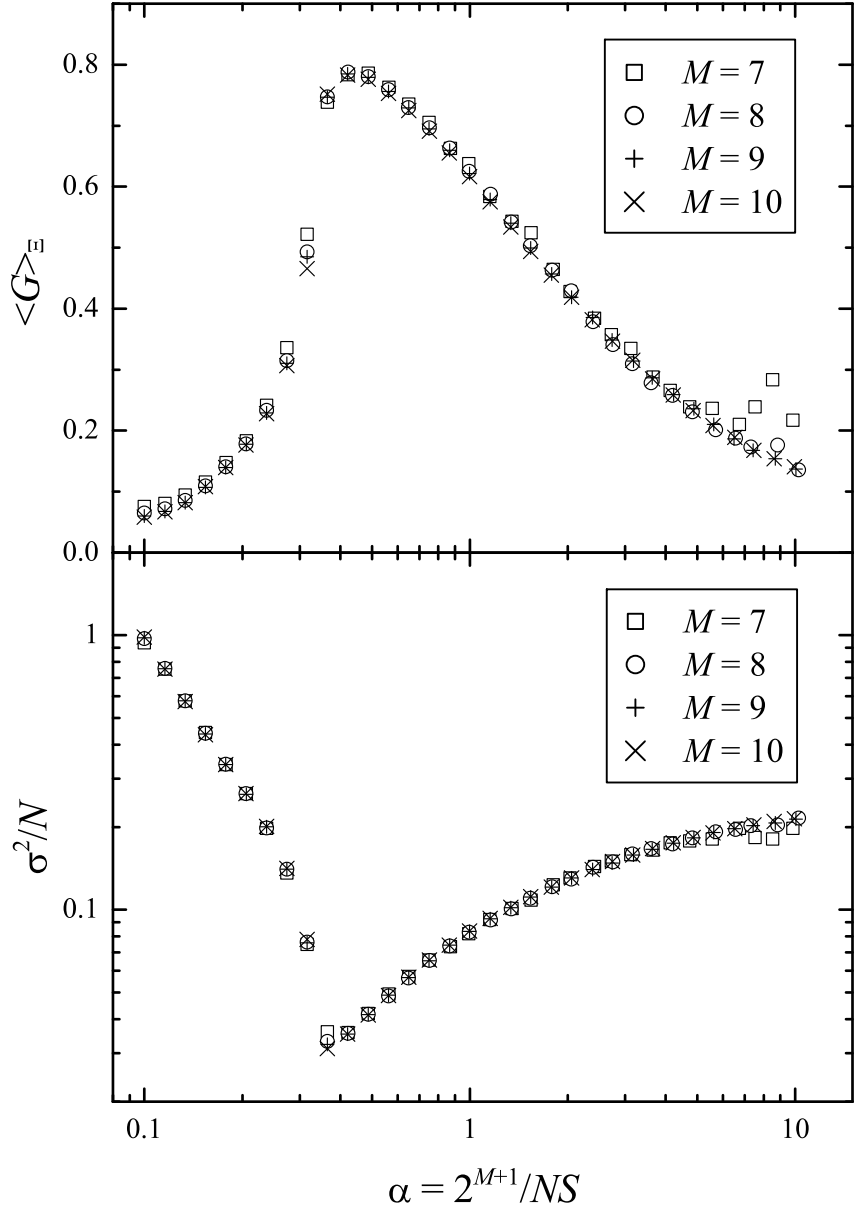}
 \caption{The Gini index $\langle G \rangle_\Xi$ and the variance of 
 attendance per player $\sigma^2 / N$ averaged over the initial configuration
 versus $\alpha$ in the MG with $S = 2$ for different values of $M$. The error
 bar of $\langle G \rangle_\Xi$ is of order of at most $10^{-3}$. The small bump 
 around $\alpha = 10$ for $M = 7$ is due to finite size effect.
}
\label{fig:f1}
\end{figure}

Let us study the Gini index averaged over the initial conditions $\langle G 
\rangle_\Xi$ versus the control parameter $\alpha$ 
as shown in Fig.~\ref{fig:f1}. (Note that we use $\langle . 
\rangle_\Xi$ to denote the average over the initial configuration of the 
system.) Our numerical results show that the curves of $\langle G 
\rangle_\Xi$ for different $M$ coincide. which means that the Gini index 
$\langle G \rangle_\Xi$, just like the variance of attendance, depends only 
on the control parameter $\alpha$ in the MG.

We now move on to discuss the properties of the Gini index $\langle G 
\rangle_\Xi$ as a function of $\alpha$ in detail. Fig.~\ref{fig:f1} shows that 
the Gini index $\langle G \rangle_\Xi$ is small when $\alpha \rightarrow 0$. 
In other words, the wealth of all players is roughly the same in such a case. 
In fact, the small value of $\langle G \rangle_\Xi$ can be explained by the 
crowd-anticrowd theory~\cite{CA1,CA2,CA3} as follows. In the small $\alpha$ 
regime, players are likely to have at least one high ranking strategy at each 
instance, as each player possesses a relatively large portion of strategies of 
the reduced strategy space. Thus, most of the players are using the crowd of 
high ranking strategies, i.e., those high ranking strategies are overcrowded. 
Due to the overcrowding of strategies, each strategy alternatively wins and 
loses one virtual score repeatedly when the same history appears, under the 
period-two dynamics \cite{MG3,P2-2}. That is to say, each strategy has 
approximately the same probability to win for any history. Therefore, all 
players have roughly the same amount of wealth and this leads to a small Gini 
index $\langle G \rangle_\Xi$. 

As the control parameter $\alpha$ increases, the Gini index $\langle G 
\rangle_\Xi$ rises rapidly and subsequently attains its maximum value when 
the number of strategies at play is approximately equal to the reduced strategy
space size. To explain this, we recall that the aim of each player in the MG is to 
maximize one's own wealth, which is achieved under the maximization of the 
global profit~\cite{MG4}. Subsequently, the attendance of each choice always 
tends to $\lfloor N/2 \rfloor$ upon equilibration for all values of $\alpha$ 
since the two alternatives are symmetric in the MG. That is to say, the system 
always ``distributes'' approximately the same amount of wealth to the 
population in each turn regardless of the value of $\alpha$.  Moreover, 
whenever $\alpha \approx \alpha_c$, unlike in the cases of symmetric and 
asymmetric phases of the MG, it is not uncommon for a player to hold only low
ranking strategies since the number of strategies at play and the 
reduced strategy space size are of the same order. Consequently, a 
significant number of players are forced to use the crowd of low ranking 
strategies and keep on losing. On the other hand, those players picking the 
crowd of high ranking strategies  have a higher winning probability and 
keep on using those strategies.  Note that the ranking of the strategies is
almost unchanged when $\alpha \approx \alpha_c$~\cite{CA1,CA2,CA3}. As a 
result, the wealth distribution of players would become relatively diverse 
and the Gini index $\langle G \rangle_\Xi$ of the population attains its 
maximum value when $\alpha \rightarrow \alpha_c$.

Actually, the increase in the Gini index when $\alpha \rightarrow \alpha_c^+$ can
be justified by the frozen probability of the MG. We recall that in the MG a player 
employs the virtual score system to determine which strategy to use in the 
next time step. In the asymmetric phase, the probability that a strategy
assigned to a player has a virtual score asymptotically higher than all 
the other strategies assigned to the same player increases as $\alpha$ 
decreases. Some players end up using only one strategy after the system
equilibrates, they are regard as frozen players. The frozen probability
indicates the number of frozen players. A small frozen probability, i.e.
most players in the game keep changing their best strategies, implies that 
only a few player will keep on winning or keep on losing all the time and
the Gini index should be low.  On the other hand, a high frozen probability 
may indicate that while some frozen players are using strategies that
win most of time, the best performing strategies for the other frozen players
are losing badly. Thus, there is a wide spread in wealth distribution of
players. The Gini index should be high in this case. The 
frozen probability follows the same trend of Gini index as $\alpha 
\rightarrow \alpha_c^+$, which further supports the validity of the result 
of the Gini index. Moreover, when $\alpha \approx \alpha_c$, it is likely that 
those frozen players which form the majority of crowds and anti-crowds in the game 
use anti-correlated strategy pairs, resulting in effective crowd-anticrowd cancellation
between frozen players. Also, those frozen players who picked the anti-correlated
strategy pairs keep winning or keep losing throughout the game.

After attaining the maximum value, the Gini index $\langle G \rangle_\Xi$ 
decreases and gradually tends to zero when the control parameter $\alpha$ 
further increases. According to crowd-anticrowd theory~\cite{CA1,CA2,CA3}, it 
is because most of the strategies at play are uncorrelated to each other when 
the strategy space size becomes much larger than the number of strategies at 
play. Therefore, it is as if each player is making random choices in the game
when $\alpha$ is large. Hence, the winning probability of all strategies is
roughly the same. As a result, the Gini index $\langle G \rangle_\Xi$ of the 
population is small in this regime.

As we have reasoned above, the winning probability of each individual player
is steady after equilibration of the system. Since the wealth distribution 
depends solely on the winning probabilities of individual players, the $g_j$'s,
and hence the Gini index $G$, converge over a sufficiently long time. Moreover, it is 
easy to check that the $g_j$'s converge logarithmically. Therefore, the 
equilibration time for $G$ is much longer than that of $\sigma^2(A)$.

\section{Semi-Analytical study of the Gini index in MG using the Replica Trick}
\label{Replica}

\subsection{Methodology}
In the previous section, we have explained the wealth inequality of 
the players in the MG qualitatively. In fact, the system of the MG can be described 
as a disorder spin system~\cite{spin1,spin2} since the dynamics of the MG indeed 
minimizes a global function related to market predictability. In
this section, we calculate the Gini index $G$ of the population in MG 
semi-analytically by mapping the MG to a spin glass. As we shall see, this 
approach works well whenever $\alpha > \alpha_c$.

Let us start to link the MG, a repeated game with $N$ players, to the spin glass. 
In this formalism, every player has to choose one out of two actions $\pm 1$ 
corresponding to the two alternatives at each time step. We denote the action 
of the $i$th player at time $t$ by $c_i(t)$. After all players have chosen their 
actions, those players choosing the minority action win and gain one unit of 
wealth while all the others lose one. In the MG, the only public information 
available to the players is the so-called history, which
is the string of the minority action of the last $M$ time steps. Namely, the 
history is a string, $(\Pi(t - M), \ldots, \Pi(t-1))$ where $\Pi(t)$ denotes 
the minority action at time $t$. For convenience, we label the history by an 
index $\mu$ as follows:
\begin{equation}
	\mu(t) = \Pi(t-M) \times 2^{M-1} + \Pi(t-M-1) \times 2^{M-2} + \ldots 
	+ \Pi(t-1).
\end{equation}
At the beginning of the game, each player picks once and for all $S$ strategies
randomly from the strategy space. In fact, a strategy specifies an action 
$a_{s,i}^{\mu}$ taken by the $i$th player for all possible histories 
$\mu = 1, \ldots,2^M$. In the MG, agents make use of the virtual score, i.e., the 
hypothetical profit for using a strategy throughout the game, to 
evaluate the performance of a strategy. To guess the next global minority 
action, each player uses their own current best strategy which is the strategy
with the highest virtual score at that moment. Assuming each player has $S = 2$
strategies which are labeled by $`+'$ and $`-'$, we define the disorder variables 
\{$\omega_i^\mu$, $\xi_i^\mu$\} as:
\begin{equation}
	w_i^\mu = \frac{a_{+,i}^{\mu} + a_{-,i}^{\mu} }{2} , \>\>
	\xi_i^\mu = \frac{a_{+,i}^{\mu} - a_{-,i}^{\mu} }{2}.
\end{equation}
Here we use the spin variable $s_i(t) = \pm 1$ to denote the strategy used 
by the $i$th player at time $t$. Thus, the action of this player is given
by
\begin{equation}
	c_i (t) = \omega_i^{\mu(t)} + s_i(t) \xi_i^{\mu(t)}.
\end{equation}
With the above formalism, we can employ a statistical tool called the replica 
trick~\cite{spin2,spin3} to study the stationary state properties of 
the MG by solving the ground state of the Hamiltonian $H$:
\begin{equation}
	H\{\vec{m}\} =  \overline{\Omega^2} + 2 \sum_{i=1}^N \overline{
	\Omega\xi_i} m_i + \sum_{i,j}^N \overline{\xi_i \xi_j} m_i m_j,
\end{equation}
where $m_i \equiv \langle s_i(t) \rangle$ and $\Omega^\mu = \sum_{i=1}^N 
\omega_i^\mu$. Note that $\overline{O}$ denotes the average over history
$\mu$ and $\langle.\rangle$ denotes the average over time $t$. In other
words, our aim is to find the minimum of $H\{\vec{m}\}$ defined by:
\begin{equation}
	\min_{\vec{m} \in [-1,1]^N} H\{\vec{m}\} = - \lim_{\beta \rightarrow 
	\infty}	\frac{1}{\beta} \langle \ln Z (\beta) \rangle_\Xi,
\end{equation}
where the partition function
\begin{equation}
	Z(\beta) = \text{Tr}_{\vec{m}}e^{- \beta H \{ \vec{m} \} }.
\end{equation}
Here, $\text{Tr}_{\vec{m}}$ denotes the integral of $\vec{m}$ on $[-1,1]^N$, 
$\langle.\rangle_\Xi$ denotes the average over the disorder variables 
$a_{s,i}^\mu$ (i.e., the quenched disorder $\Xi$ of the system) and $\beta$ 
stands for the inverse temperature. In fact, the ground state solution of $H$ 
depends on the disorder variables. However, in the thermodynamic limit, the
ground state of $H$ has a unique solution for all quenched disorder. Thus, 
in the replica calculation, we seek for ground state solution of the Hamiltonian
$H$ on average of the quenched disorder. In order to evaluate $\langle \ln Z 
\rangle_\Xi$, we construct the partition function $Z^n$ by studying $n$ (a 
non-negative integer) replicas of the system with identical disorder variables 
$\{a_{s,i}^\mu\}$. Then, we perform a semi-analytical continuation to extend this 
function for non-integer $n$. In this way, the average of $\ln Z$ over
$\{a_{s,i}^\mu\}$ is reduced to
\begin{equation}
	\langle \ln Z \rangle_\Xi = \lim_{n \rightarrow 0}\frac{1}{n} \ln 
	\langle Z^n \rangle_\Xi.
\end{equation}
We also define the free energy density $F_\beta(\hat{Q},\hat{r})$ by
\begin{equation}
	\langle Z^n \rangle_\Xi = \int d \hat{r} \int d \hat{Q}
	\exp\left[ - \beta n N F_\beta(\hat{Q}, \hat{r}) \right],
\end{equation}
where $Q_{a,b} = \frac{1}{N} \sum_i^N m_i^a m_i^b$ is the overlap matrix and
$r_{a,b}$ are the associated Lagrange multipliers. Hence, we can find the 
stationary state solution of $H$ in the thermodynamic limit $N \rightarrow
\infty$ by finding the minima of $F_\beta(\hat{Q},\hat{r})$ as: 
\begin{equation}
	\lim_{N \rightarrow \infty} \min_{\vec{m} \in [1,-1]^N} \frac{H\{m\}}{N}
	\approx \lim_{\beta \rightarrow \infty} \lim_{n \rightarrow 0} \min 
	F_\beta(\hat{Q},\hat{r}).
\end{equation}

In fact, we can find the minima in the replica symmetric ($RS$) ansatz by 
solving the saddle point equations~\cite{spin3,MMG1}:
\begin{equation}
	\frac{\partial F_\beta}{\partial r_{a,b}} = 0 \>\> \text{and} \>\>
	\frac{\partial F_\beta}{\partial Q_{a,b}} = 0 \>\> \forall a,b .
\end{equation}
In this ansatz, the matrices $\hat{r}$, $\hat{Q}$ corresponding to
$\min F_\beta$, are assumed to be in the following form:
\begin{equation}
	Q_{a,b} = \frac{1}{N}\sum_i^N m_i^a m_i^b = \left\{ \begin{array}{ll}
                                                q & \mbox{for $a \neq b$}\> ,\\
                                                Q & \mbox{for $a = b$} \> ,
                                                \end{array}
                                            \right. 
\end{equation} 
and
\begin{equation}
r_{a,b} =  \left\{ \begin{array}{ll}
                        2r & \mbox{for $a \neq b$} \>,\\
			R & \mbox{for $a = b$} \> ,
                     \end{array}
            \right. 
\end{equation}
for all $a,b=1,2,\ldots ,n$.
Therefore, using the $RS$ ansatz, the minimum value of $F_\beta$ in the $n 
\rightarrow 0$ limit is given by
\begin{eqnarray}
	F^{(RS)} &=& \lim_{n \rightarrow 0} \min F_\beta (\hat{Q}, \hat{r}) 
	\nonumber \\
	&=& \frac{\alpha}{2 \beta} \ln \left[ 1 + \frac{\beta}{\alpha}
	(Q - q)\right] + \frac{\alpha (1 + q)}{2 [\alpha + \beta ( Q - q)]}
	- \frac{1}{\beta} \int d \Phi(\lambda) \ln \left[ \int_{-1}^{1} d m 
	\exp( - \beta V(m|\lambda)) \right] + \frac{\alpha \beta}{2} ( RQ - rq),
	\nonumber \\
\end{eqnarray}
where $\Phi(\lambda)$ is the normal distribution and the potential 
$V(m|\lambda) = -\sqrt{\alpha r}\lambda m + \frac{\alpha \beta}{2}
(r - R)m^2$. 

Using the saddle point equations, we arrive at~\cite{spin3,MMG1}
\begin{equation}\label{rho}
\frac{\alpha}{\rho^2} = 2 - \sqrt{ \frac{2}{\pi} } \frac{1}{\rho}e^
{-\frac{\rho^2}{2}} - \left( 1 - \frac{1}{\rho^2} \right)\text{erf} \left( 
\frac{\rho}{\sqrt{2}}\right),
\end{equation}
where $\rho$ is a disorder variable and depends on the control parameter 
$\alpha$. The probability distribution of the `average action' of a player,
$m$, is then given by~\cite{spin2}
\begin{equation}
P(m) = \frac{\phi(\rho)}{2} [ \delta(m-1) + \delta(m+1)] + 
\frac{\rho}{\sqrt{2 \pi }} e^{ - (\rho m )^2 / 2}, \label{m_dis}
\end{equation}
where $\phi(\rho) = 1 - \text{erf} ( \rho / \sqrt{2} )$, $\delta(0) = 1$ 
and $\delta(x) = 0$ whenever $x \neq 0$. Note that Eqs.~(\ref{rho})
and~(\ref{m_dis}) are only valid for $\alpha > \alpha_c$. For $\alpha < 
\alpha_c$, the replica calculation cannot give correct predictions for the 
probability distribution of spin variable $m$ because it is unable to 
reproduce the period-two dynamics of the system~\cite{spin3}. 

Our aim is to calculate the Gini index of the players in the MG 
using the replica trick. At first glance, one might argue that 
the distribution of $g_i$ can be reproduced analytically using the replica trick.
However, the replica trick can only generate the average gain of a group of 
players rather than the wealth of an individual player. This is why Challet
did not compute the theoretical gain of individual players analytically
by using the replica trick for the MG. In fact, he computes the gain 
semi-analytically using the disorder spin variable $m_i$ measured in the simulations
instead~\cite{MMG1,PC}.  To reproduce the wealth distribution of players, 
we need to know the actions of each individual player $s_i(t)$ at time $t$ for 
each individual particular quenched disorder. However, $s_i(t)$ cannot be found 
by the replica trick. So we approximate $s_i(t)$ by the disorder spin 
variable $m_i$ generated stochastically from the distribution $P(m)$ which 
is found by the replica trick. Then, the Gini index $G(\Xi)$ can be calculated 
from the wealth distribution of the players for that quenched disorder. 
As we are only interested in the generic properties of the Gini index, 
we calculate the Gini index averaged over quenched disorder $\langle G 
\rangle_\Xi$. This should be done by calculating the Gini index of each individual 
quenched disorder $G(\Xi)$ first and then taking average over all quenched disorders.

In practice, we perform the stochastic simulation to generate the wealth
distribution of population in the MG for an individual quenched disorder in the 
following way. Before starting the simulation, the quenched
disorder $\Xi$ is formed by allowing each player to pick two strategies 
randomly from the reduced strategy space. Next, each player draws the 
spin variable $m$ from the distribution $P(m)$ as shown in Eq.~(\ref{m_dis}). 
Those players with $m = \pm 1$ are called frozen players because they 
keep on using a strategy throughout the game. Then, in each step of the game, 
players choose one of their own strategies according to their own spin
variable $m$ to guess the next global minority side. In practice, 
the strategy used by the $i$th player at time $t$, $\sigma_i(t)$, is chosen by 
calling a uniform random variate $\zeta$ on $[-1,1]$. Then we set $\sigma_i(t) 
= 1$ if $m_i \geq \zeta$ and $\sigma_i(t) = -1$ otherwise. Therefore, for the 
history $\nu(t)$, the action of the $i$th player at time $t$ can be written as
\begin{equation}
	\chi_i(t) = \omega_i^{\nu(t)} + \sigma_i(t) \xi_i^{\nu(t)}. 
	\label{action}
\end{equation}
Note that the history $\nu(t)$ is generated randomly at each time step $t$ in 
our simulation. In addition, the difference in the numbers of players choosing 
the two alternatives at time $t$ is given by
\begin{equation}
	X(t) = \sum_{ i = 1 }^N \chi_{i}(t).
\end{equation}
So we obtain the minority side at time $t$
\begin{equation}
	\Theta(t) = -\text{sign}( X(t) ).
\end{equation}
After determining the minority side, the wealth of the $i$th players, $w_i(t)$, 
is updated by
\begin{equation}
	w_i(t+1) = w_i(t) + 2\delta(\chi_i(t) - \Theta(t)) -1. \label{wealth_up}
\end{equation}

We repeat the above algorithm $N_s$ times for the system to equilibrate.
After the equilibration, we measure on the Gini index of the 
population for the quenched disorder $\Xi$ using Eq.~(\ref{gini_eq}). Then
we calculate the average Gini index for $500$ independent runs. We denote the Gini index 
calculated by this algorithm with averaging over the quenched disorder by 
$\langle G \rangle_\Xi^{\text{R}}$. In fact, we find that the average Gini 
index $\langle G \rangle_\Xi$ converges after $N_s = 500 P$ iterations,  
where $P = 2^M$ is the number of possible histories.

\subsection{Semi-Analytical Results Using Stochastic Simulation}

\begin{figure}[ht]
\includegraphics*[scale = 0.4]{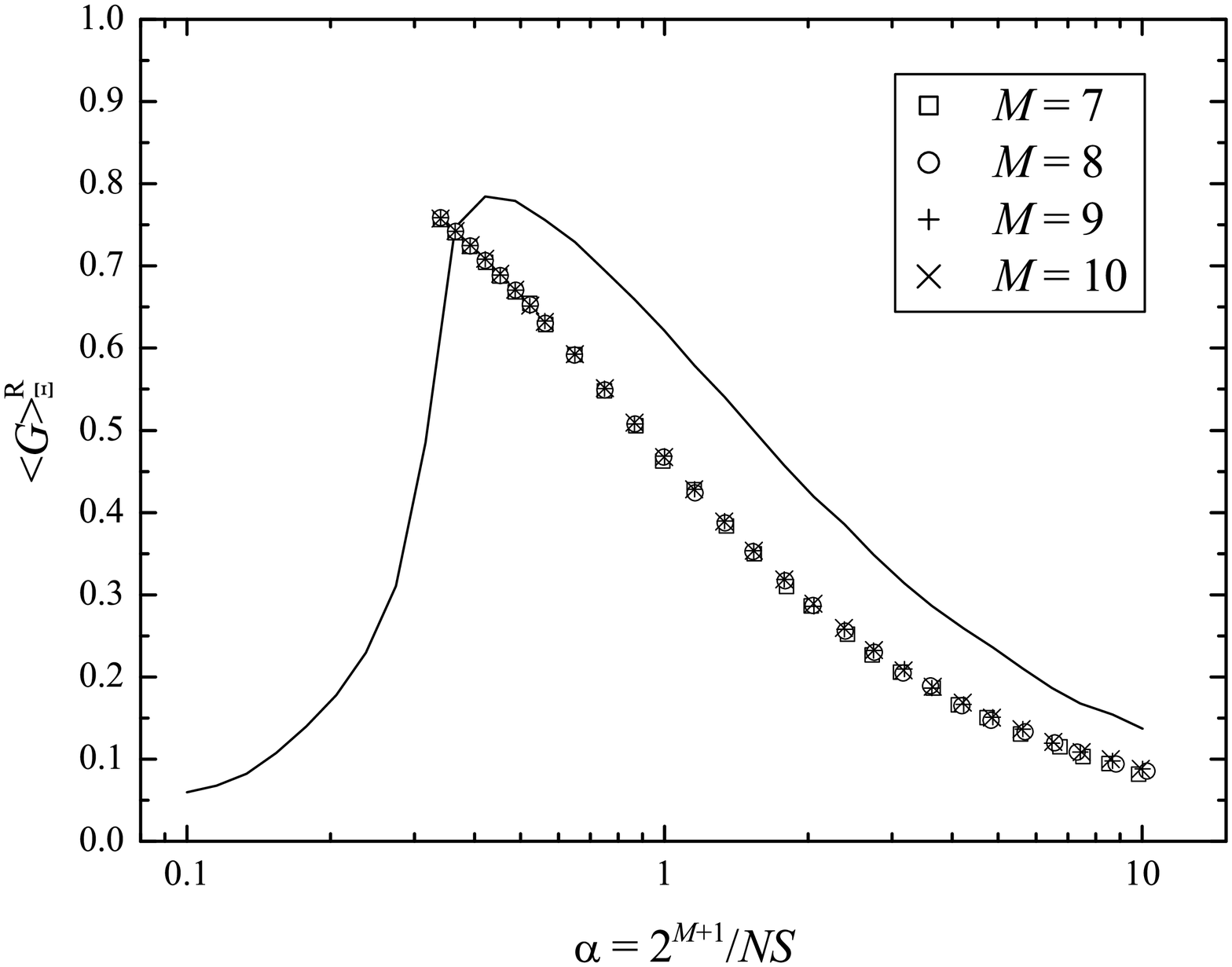}
 \caption{The average Gini index found in stochastic simulation using random 
 history $\langle G \rangle_\Xi^{\text{R}}$ versus the control parameter
 $\alpha$ in the asymmetric phase of the MG with $N_s = 500 P$ and $S = 2$ for 
 different $M$. For comparison purpose, the solid line indicates the 
 corresponding numerical results in the MG with $M = 9$.}
\label{fig:f2}
\end{figure}

Fig.~\ref{fig:f2} gives the Gini index obtained from semi-analytical calculation of
$\langle G \rangle_\Xi^{\text{R}}$ versus the control parameter $\alpha$ for 
MG with $\alpha > \alpha_c$. We find that the trend of the curves of $\langle 
G \rangle_\Xi ^{\text{R}}$ agrees with the numerical findings. This implies that 
we have successfully reproduced the numerical results of the Gini index in the 
asymmetric phase of the MG by using the replica method. However, the curves of 
$\langle G \rangle_\Xi^{\text{R}}$ are systematically lower than those from 
numerical simulation. This is because the coupling between the actions of 
players and the dynamics of the system is completely ignored in our stochastic 
simulation as the actions of the players depend only on the spin
variable $m$. Consequently, the global cooperation among the players is 
suppressed in our semi-analytical calculation. Hence, the wealth distribution of 
players is less diverse which results in under-estimation of the Gini index in 
the MG.

\begin{figure}[hb]
\includegraphics*[scale = 0.4]{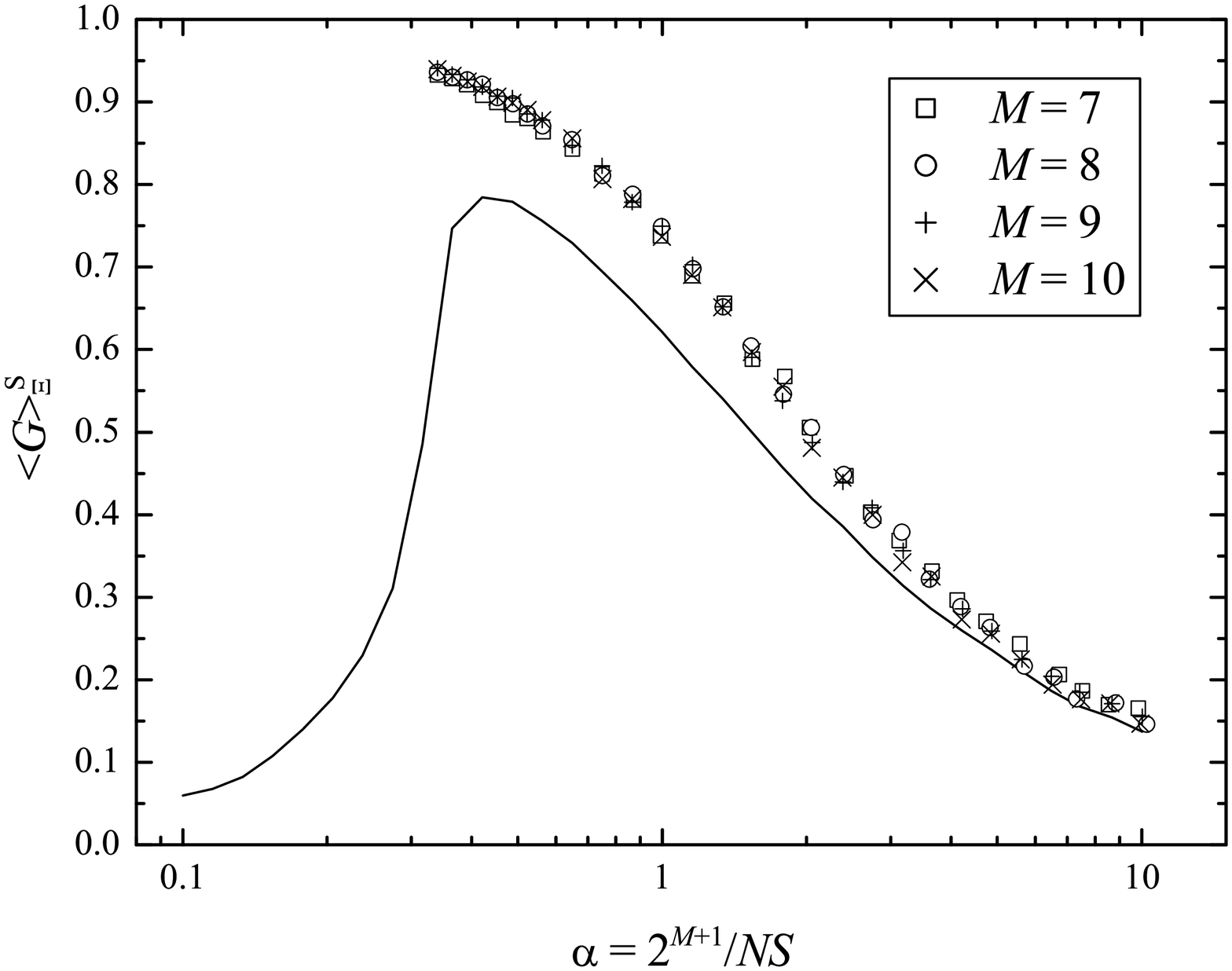}
  \caption{The average Gini index found in stochastic simulation using 
 sequential history $\langle G \rangle_\Xi^{\text{S}}$ versus the control 
 parameter $\alpha$ in the asymmetric phase of MG with $N_s = 500 P$ and 
 $S = 2$ for different $M$. For comparison purpose, the solid line indicates 
 the corresponding numerical results in MG with $M = 9$.}
\label{fig:f3}
\end{figure}

To make our semi-analytical calculation more ``realistic'', we allow the
history $\nu(t)$ to be updated sequentially by
\begin{equation}
	\nu(t) = [ 2 \nu(t-1) + \Theta(t) ] \bmod P ,
\end{equation}
and we denote the Gini index averaged over the quenched disorder calculated in this 
approach by $\langle G \rangle_\Xi^{\text{S}}$. Note that 
$\langle G \rangle_\Xi^{\text{R}}$ and $\langle G \rangle_\Xi^{\text{S}}$ 
are calculated using the same algorithm except that the history is updated 
in a different way. Fig.~\ref{fig:f3} shows the Gini index $\langle G 
\rangle_\Xi^{\text{S}}$ versus the control parameter $\alpha$ in the MG. We 
observe that the values of $\langle G \rangle_\Xi^{\text{S}}$ agree well with 
the numerical results when $\alpha$ is large. According to crowd-anticrowd 
theory, if $\alpha$ is large, most strategies of the players are uncorrelated 
to each other due to the under-sampling of the strategy space. Moreover, most 
of the strategies are used by either one or none of the players in the MG whenever 
$\alpha \rightarrow \infty$. Therefore, the cooperation between the players can
be neglected for $\alpha \rightarrow \infty$. In addition, the probability of 
the occurrence of different histories is not the same in the MG when $\alpha 
\rightarrow \infty$~\cite{Mem}. Indeed, these two conditions are satisfied in 
our stochastic simulation using the sequential history. So, the values 
of $\langle G \rangle_\Xi^{\text{S}}$ match the numerical estimates when $\alpha$ is large. 

On the other hand, when $\alpha$ approaches $\alpha_c^+$, the values 
of $\langle G \rangle_\Xi^{\text{S}}$ become larger than the 
numerical results. This discrepancy can be explained as follows. As mentioned
in Section~\ref{Num_Re}, since there is effective crowd-anticrowd cancellation,
the history in the MG becomes more uniform as $\alpha$ approaches $\alpha_c^+$~\cite{Mem}.
In contrast, although players still have the same chance to pick
anti-correlated pairs separately at the beginning of the game in our
sequential simulation, the strategy actually used by each player at each turn is
not determined by its virtual score, but a randomly assigned
disorder spin variable $m_i$ instead. Consequently, two players are less 
likely to be frozen on an anti-correlated strategy pair. This makes the 
crowd-anticrowd cancellation less effective among frozen players in our 
sequential simulation. So, the actions among these frozen players may give a 
strong bias in the output, especially for $\alpha \approx \alpha_c$, where 
frozen probability is highest. In turn, the history becomes much more 
non-uniform. This greatly increases the Gini index as some players have more 
chance to stay at the winning (or losing) side.

Finally, we remark that both $\langle G \rangle_\Xi^{\text{R}}$ and $\langle G 
\rangle_\Xi^{\text{S}}$ calculated by the stochastic simulation are independent
of $M$. This is expected, since the results of the replica calculation do not 
depend explicitly on $M$. 

\section{Conclusion} \label{Con}

In summary, we have investigated the inequality of wealth among players 
in the MG using the well-known measure in economics called the Gini index. In 
particular, our numerical findings show that the wealth inequality of players 
is very severe near the point of maximum global cooperation $\alpha_c$. That is
to say, in the minority game, global cooperation comes hand in hand with
uneven distribution of players' wealth. Specifically, a significant number of 
players are forced to use the low ranking strategies and cooperate with those 
players using the high ranking strategies since the number of strategies at play 
and the reduced strategy space size are of the same order whenever $\alpha \rightarrow 
\alpha_c$. In this respect, we have showed that the crowd-anticrowd theory 
offers a simple and effective platform to study the wealth inequality in the MG.

In addition, we have studied the Gini index semi-analytically by mapping the 
system of the MG to a spin glass. With this formalism, we semi-analytically reproduce 
our numerically simulated Gini index in the asymmetric phase of MG by 
investigating the stationary state properties of MG using the replica trick.

\section{Acknowledgement}
We would like to thank the Computer Center of HKU for their helpful 
support in providing the use of the High Performance Computing Cluster 
for the simulation reported in this paper. Useful conversations with
W.~C. Man are also gratefully acknowledged.

\end{document}